\documentclass[%
notitlepage,
onecolumn,
nofootinbib,
amsmath,amssymb,
aps,
pra,
]{revtex4-2}

\usepackage{graphicx}
\usepackage{physics}
\usepackage{dcolumn}
\usepackage{bm}
\usepackage{soul}

\usepackage{color}

\usepackage{soul}
\usepackage{xcolor}
\definecolor{col}{rgb}{0.0,0.0,0.0} 
\usepackage{ulem}

\begin{document}

\title{Multifaceted phase ordering kinetics of an~antiferromagnetic spin-1 condensate}

\author{Joanna Pietraszewicz}
\affiliation{Institute of Physics, Polish Academy of Sciences, Aleja Lotnik\'ow 32/46, PL-02668 Warsaw, Poland}

\author{Aleksandra Seweryn}
\affiliation{Institute of Physics, Polish Academy of Sciences, Aleja Lotnik\'ow 32/46, PL-02668 Warsaw, Poland}

\author{Emilia Witkowska}
\affiliation{Institute of Physics, Polish Academy of Sciences, Aleja Lotnik\'ow 32/46, PL-02668 Warsaw, Poland}

\begin{abstract}
We study phase domain coarsening in the long time limit after a quench of magnetic field in a~quasi one-dimensional spin-1 antiferromagnetic condensate. We~observe that the~growth of correlation length obeys scaling laws predicted by the two different models of phase ordering kinetics, namely the~binary mixture and vector field. We derive regimes of clear realization for both of them. We~demonstrate appearance of atypical scaling laws, which emerge in~intermediate regions.
\end{abstract}

\maketitle

\section*{Introduction}
\label{sec:intro}

The theory of phase ordering kinetics~(PhOK) states that the growth of order occurs through the coarsening of phase domains, when the system is quenched from disordered to ordered phase. 
The typical length scale of the phase domain increases then with time. According to the dynamic scaling hypothesis it is due to a global change of scale. In a~homogeneous system, the typical size of a~phase domain is characterized by the correlation length $L(t)$, which can be defined as the half-width of the~equal-time correlation function for the local order parameter $\phi(x,t)$,
\begin{equation}\label{eq:correlation_function}
g^{(1)}(x, t)=\Big\langle\, {\int {\rm d}x'\  \phi(x'+x,t)^* \phi(x',t)\,} \Big\rangle .
\end{equation}
The angle brackets in Eq.(\ref{eq:correlation_function}) indicate the average over initial conditions representing a~disordered state. These initial conditions defined in the momentum representation are such $\langle \phi(k,0)^* \phi(k',0) \rangle = \Delta \, \delta(k-k')$, where $\Delta$~sets the~size of initial fluctuations in the field~$\phi$. The scaling hypothesis implies that due to the existence of the unique characteristic length scale $L(t)$, the correlation function is just one-parameter dependent, i.e. $g^{(1)}(x, t) = f\left(x/L(t) \right)$.
Therefore, the central interest of the~theory is in the time evolution of the correlation length, and whether this function follows universal scaling laws $L(t)\sim t^{1/z_d}$. Further, if yes, what is the value~of~the dynamical exponent $z_d$. The importance attributed to $z_d$ stems from its universal character. Since the~exponent does not depend on microscopic properties, it reveals general features~of an~entire class of systems~\cite{Bray2002}.

Kinetic models for the derivation of particular scaling functions were extensively studied and established in 1990s. The~models were devoted to classical systems, and the~general framework associated with the Hohenberg and Halperin A-J classification of dynamic critical phenomena~\cite{RevModPhys.49.435} made the~theory more universal.
Several systems belonging to the~same dynamical class of models exhibit the same universal scaling laws determined by the physical mechanism embedded into them. The phase ordering kinetics, for example, is~governed by the H model for binary liquids, in which hydrodynamic processes dominates. On~the~other hand, the B model associated with vector fields is controlled by diffusion processes mainly. Even though both models are conservative, the scaling exponents corresponding to each of them are different. Nowadays, the~subject of the PhOK has been revived in quantum systems, a good example of which are ultra-cold atomic gases \cite{Bookjans11,Erne18,Johnstone19,calabrese05,tkachev97,sciola13}. In particular advanced studies concern nonthermal fixed points \cite{orioli15,Karl17,Mikheev19,Chantesana19,Gasenzer19}. The vast majority of works is devoted to two- and three-dimensional spin-1 ferromagnetic condensates \cite{PhysRevLett.116.025301, PhysRevLett.119.255301, PhysRevA.96.013602, PhysRevA.95.023616, 1367-2630-19-9-095003, PhysRevA.94.023608, PhysRevA.95.053638, PhysRevA.91.053609, PhysRevA.88.013630, Oberthaler2018,Symes18,Schmied19,lamacraft07,10.21468/SciPostPhys.7.3.029}, but also for one-dimensional spinor~\cite{Vinit13, Fujimto19, Fujimoto18} or binary Bose condensates \cite{PhysRevLett.113.095702,PhysRevA.92.043608,torre13}, including driven-dissipative systems \cite{Comaron18,PhysRevB.95.075306,Kagan94}. 

In this paper the PhOK is investigated in a quasi one-dimensional antiferromagnetic spin-1 condensate. The sudden quench of a weak magnetic field leads to the transition from the antiferromagnetic state to a state where domains of atoms with different spin projections separate. We are particularly interested in the superfluid order for atoms in the $m_F=0$ Zeeman state. We~have performed numerical calculations within the truncated Wigner approximation~\cite{Sinatra_2002} and made the following observations. The scaling hypothesis holds for any regime of parameters, however, on the longest time scale, the whole variety of scaling exponents is observed. Their values are~$3$ or~$3/2$ both in the sense of hydrodynamic (H) models~\cite{PhysRevLett.116.025301, PhysRevLett.119.255301, PhysRevA.95.023616, PhysRevA.91.053609,PhysRevA.88.013630}, and even~$4$ (with a logarithmic correction) in the sense of the vector's field (B) model --- another class of the model of the Hohenberg and Halperin classification. All this occurs depending on the system's parameters. The scaling exponent $z_d=4$ has not yet been reported in the PhOK studies with quantum systems.

Using parameters for which we observe a change of scaling exponent passing from the short to long time scales, the characteristic length~$L(t)$ can be even a subject to multi-scaling behaviour~\mbox{\cite{Castellano96, Coniglio_1989, Fujimoto18}}. The reason for such a~multifaceted PhOK is that \mbox{the spin-1} antiferromagnetic condensate effectively behaves as a~binary mixture or vector fields model which is assigned to the~H or B model, respectively. Therefore, the~PhOK of the system can exhibit features characteristic for the~H or B models independently, or both of them simultaneously. A~study of the~interplay between the models can be made. This is interesting because to date studies of the~PhOK attributed system's behaviour mostly to a particular scaling law characterised by a particular model. A natural questions arises  whether distinct models, and hence their physical mechanisms, are mutually compatible, or under what conditions they co-occur. Here, we~characterize and classify the appearance of various scaling exponents. We~calculate borders for the~limit cases in which the~B and H models can be realized in their forms. In~the~region around borders, both models compete, leading to various scaling laws in which scaling exponents smoothly change among the~two limiting cases.  

\section{Model and methods}

\begin{figure}[h!]
\centering
\includegraphics[width=0.9\linewidth]{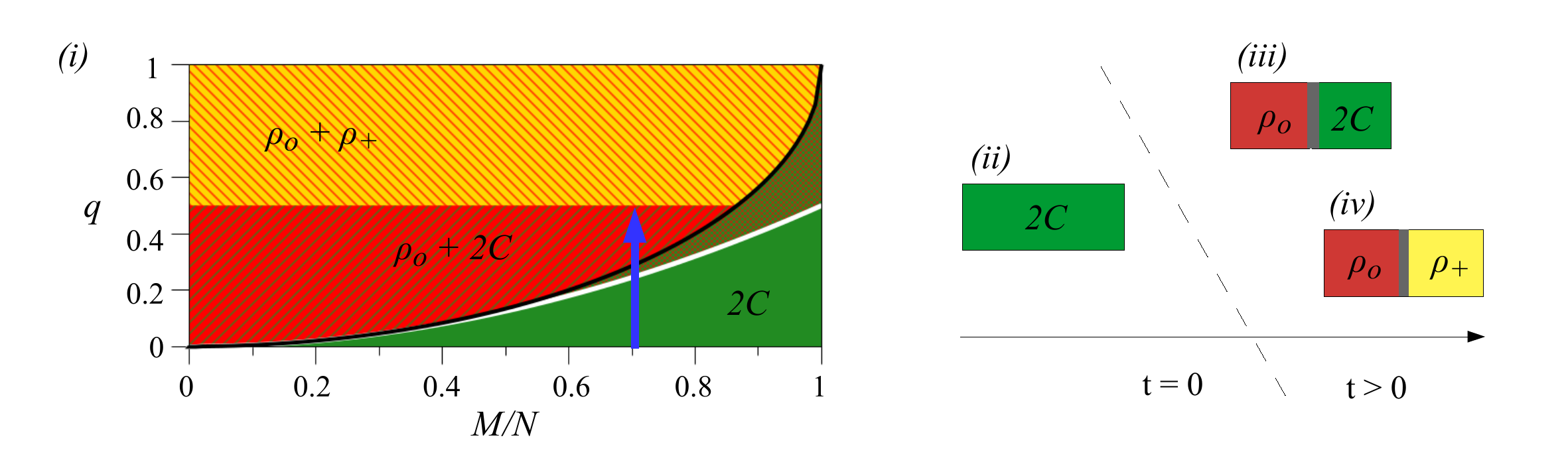}
\caption{$(i)$ The phase diagram of antiferromagnetic BECs~\cite{Matusz09,Nature2019Gerbier} in the thermodynamic limit (when the spin healing length $\xi_s= \hbar/\sqrt{2m\, c_2\rho}$ is much smaller than the linear system size $L$, i.e., $\xi_s \ll L$) and positive magnetization. Three homogeneous phases (marked by colors) can be revealed: \emph{2C phase} (green) in which the~components $m_F=\pm 1$ coexist; \emph{$\rho_{0}$}~(red) and \emph{$\rho_{+}$} (yellow) \emph{phases} in which~atoms occupy the $m_F=0$ and $m_F=+1$ Zeeman state, respectively. The ground state of the system is $(ii)$ the $2C$ phase when $q<q_1=(M/N)^2/2$ (solid white line), $(iii)$ the phase separated into $2C$ and $\rho_0$ domains for $q\in(q_1,q_2)$,\linebreak and $(iv)$~the phase separated into $\rho_+$ and $\rho_0$ domains when $q\gg q_2$. The \emph{2C} phase remains a local energy minimum up to $q_c=1-\sqrt{1-(M/N)^2}$ (solid black line)~\cite{PhysRevB.88.054508}. $q_2$ is an estimation of a~crossover between the $\rho_0+2C$ and $\rho_++\rho_0$ phases~\cite{Matusz09}. Its exact value depends on $M/N$ (see~Fig.~\ref{fig:F6}). The~vertical thick lines in $(iii)$ and $(iv)$ illustrate domain walls. The~blue arrow in $(i)$ shows an example of the considered quench for $M/N=0.7$ and $q=0.5$.}
\label{fig:F1}
\end{figure}

A spinor Bose-Einstein condensate of $N$ sodium atoms is considered \cite{KAWAGUCHI2012253,Nature2019Gerbier}. The system is represented by the~vector $\vec{\psi} =(\psi_1,\psi_0,\psi_{-1})^T$, whose components describe atoms in the~corresponding Zeeman levels numerated by the magnetic number \mbox{$m_F=0,\pm 1$.} We~assume \mbox{a ring-shaped} quasi-one-dimensional geometry with periodic boundary conditions~\cite{PhysRevLett.110.025301, PhysRevLett.119.190403, Wolf2019}, where transverse degrees of freedom are confined in a strong potential with frequency~$\omega_\perp$. The Hamiltonian of the system is
\begin{equation} \label{En}
H = \int {\rm d}x \left[ \vec{\psi}^{\, \dagger} \left( -\frac{\hbar^2}{2 m} \nabla^{2}+q\, c_2 \rho\,  f_z^2\right) \vec{\psi} + \frac{c_0}{2} n^2 +  \frac{c_2}{2}\,  {\bf F}^2 \right] ,
\end{equation}
where $m$ is the atomic mass, 
$n=\sum n_{m_F}=\sum\psi_{m_F}^\dagger \psi_{m_F}$ is the local atom density and ${\bf F}=(\psi^{\dagger}f_x\psi,\psi^{\dagger}f_y\psi,\psi^{\dagger}f_z\psi)$ is the spin density with the spin-1 matrices $f_{x,y,z}$. The~spin-independent and spin-dependent interaction coefficients, $c_0$ and $c_2$, are both positive for~sodium atoms. Namely, they are $c_0=2 \hbar \omega_\perp\, (2 a_2 + a_0)/3$ and $c_2= 2 \hbar \omega_\perp\, (a_2 - a_0)/3$, where $a_S$ is the $s$-wave scattering length for pairs of colliding atoms with total spin $S$~\cite{KAWAGUCHI2012253}. 
The~term $q\, c_2 \rho$ is the quadratic Zeeman energy, where the dimensionless parameter $q$ can be controlled using magnetic field or the microwave dressing~\cite{KAWAGUCHI2012253}, and $\rho$ is the mean density of the~system. The Hamiltonian conserves the total atom number $N=\sum_{m_F}N_{m_F}$ and the~magnetization $M=N_1-N_{-1}$. The ground state of the system in the thermodynamic limit~\cite{Matusz09} is presented in~Fig.~\ref{fig:F1}. 

In the present study we consider the~non-negative magnetization and choose the $2C$ state as an initial state. Next, the~quadratic Zeeman shift is set to a fixed value $q>q_c$, and the~evolution starts. The~reason to consider quench from $2C$ towards $2C+\rho_0$, $\rho_++\rho_0$ phases is that the symmetry breaking occurs in this direction. We describe dynamics of the system on the mean-field level by solving the time-dependent Gross-Pitaevskii (GP) equations
\begin{align}
 i \hbar\, \dot{\psi}_{m_F}=&\left[ 
 -\frac{\hbar^2\nabla^2}{2m} 
 + c_0 n + c_2 (n_1-n_{-1})\, m_F 
 + c_2 (n_1 + n_{-1})\, \delta_{m_F, 0}
 + c_2 n_0\, |m_F|\right] \psi_{m_F} 
 \nonumber\\ 
 & \quad+  q\, c_2 \rho \, |m_F|\, \psi_{m_F} + c_2\, |m_F|\, \psi^*_{-m_F}\psi_0^2 + 2 c_2\, \delta_{m_F, 0}\ \psi_0^*\, \psi_1\psi_{-1},
 \label{eq:GPEs}
\end{align}
where $n_{m_F}=|\psi_{m_F}|^2$ and $\delta_{m_F,0}$ is the Kronecker delta function, see~e.g.~\cite{RevModPhys.85.1191}. To obtain the initial state for an arbitrary chosen value of $M$, all atoms are prepared in the polar ground state $\vec{\psi}_{\rm pgs}=(0,\psi_{\rm pgs}(x),0)^T$. This~state is then subject to double rotations: \mbox{($i$) a spin-1} rotation $e^{ i f_y \pi/2}$, which produces the~intermediate state $\frac{\psi_{\rm pgs}}{\sqrt{2}}(1,0,-1)^T$, and ($ii$) a rotation $e^{-i\sigma_{y} \theta}$ through angle~$\theta$ that is performed on the $m_F=\pm 1$ levels around \mbox{the~y-Pauli} matrix $\sigma_y$. The above procedure leads to \mbox{$\vec{\psi}_M=\frac{\psi_{\rm pgs}}{\sqrt{2}}(\sin\theta+\cos\theta,0,\sin\theta-\cos\theta)^T$,} and the~desired state for a given $M$ is constructed when $2\theta~=~\arcsin{\left(\frac{M}{N}\right)}$. The~state for arbitrary~$M$ can be also prepared experimentally by applying the two subsequent electromagnetic pulses~\cite{phdHamley}. In our calculations, stochastic white noise with variance \mbox{$\Delta~=~\frac{1}{2}\frac{\text{particle}}{\text{momentum mode}}$} is added to all Zeeman components of the initial $\vec{\psi}_M$ to seed the formation of symmetry-breaking phase domains. The~calculations are made for an~ensemble of $N_r=300$ realizations, the~number of grid points $\mathcal{N}=2^{10}, 2^{11}$ on the~box size $L$ large enough to avoid finite size effects, i.e., $L=7\times 10^2$, $10^3$, $5\times 10^3$, $10^4$~$\mu$m, which corresponds to $\rho=14.3, 10, 2, 1$~$\mu$m${}^{-1}$, respectively. We set the~number of~atoms to $N=10^4$.

\begin{figure}[htb!]
\centering
\includegraphics[width=0.6\textwidth]{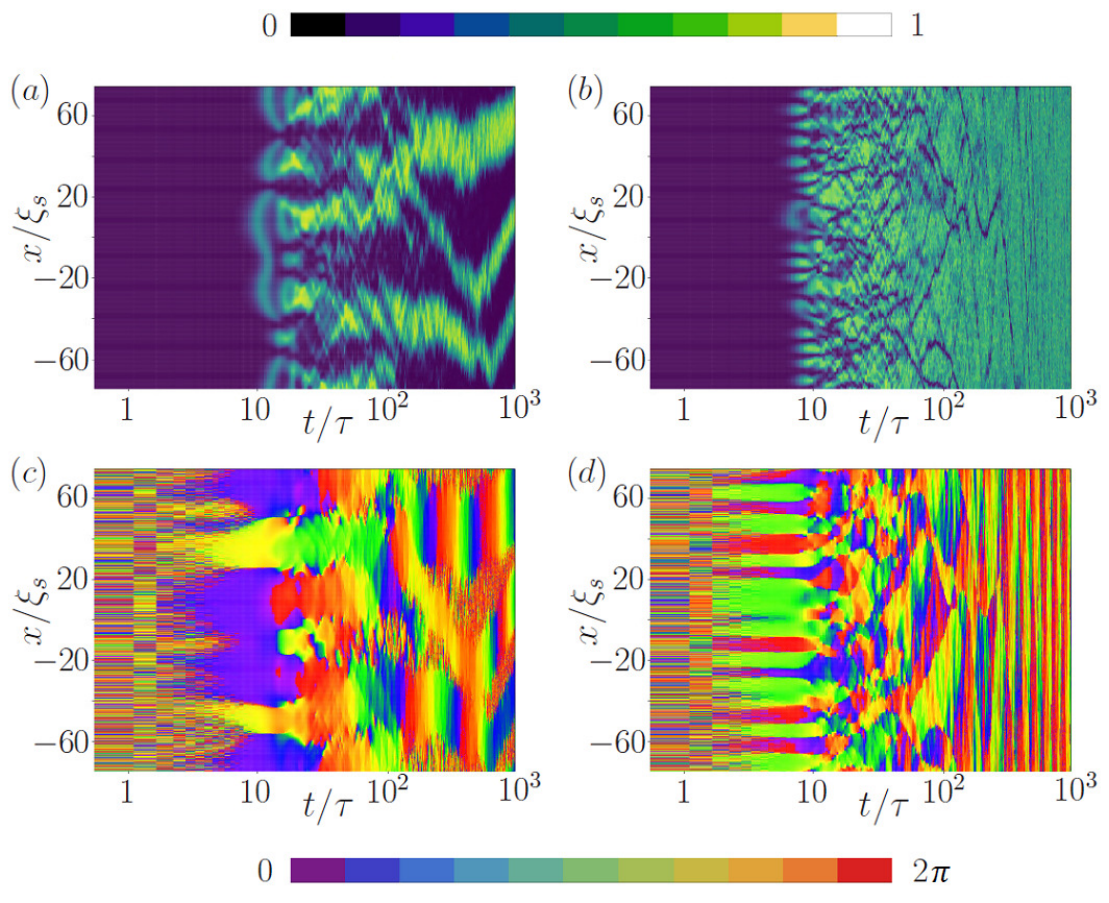}
\caption{Evolution of the normalized density (a, b) and the phase (c, d) of the wave function $\psi_0(x,t)$ describing atoms in the~$m_F=0$ Zeeman component, where $N=10^4$, \mbox{$\omega_\perp=1000$Hz,} \mbox{$\rho=14.3\mu$m$^{-1}$,} $\xi_s=9.3\mu$m and $\tau=63.2$ms. $(M,q)=(N/2,0.5)$ (left column), and \mbox{$(M,q)=(0,1.2)$} (right column). 
}
\label{fig:F2}
\end{figure}

As~illustrated in Fig.~\ref{fig:F2} in the example for the $m_F=0$ component, the growth of phase domains and their coarsening is observed. The evolution of the density and phase can be divided into three stages: (i) creation of domains seed followed by spin domain formation \mbox{around~$t=10\tau$,} (ii)~early dynamics characterized by fast reduction of the number of domains, and (iii)~further dynamics leading to domains merging at~the~longest time scale.

The~initial exponential growth followed by the phase domains formation is well understood~\cite{PhysRevB.88.054508} so far relying on unstable modes. Here, in turn, PhOK in the long time limit driven by phase domains merging is described. The description relies on the evolution~of the correlation length determined from the first-order correlation function $g^{(1)}_{N} (x, t)=\langle \int {\rm d}x'\ \psi_0(x'+x,t)^* \psi_0(x'+y,t) \rangle$, where $\psi_0(x,t)$ is a solution of GPEs for the $m_F=0$ Zeeman component. The~computed correlation length~$l_h$ has such a property that $g^{(1)}_N(l_h,t)=h$ (in~this paper $h=\frac{1}{2}$). Our initial condition, although trivialized, determines the following. No~atoms in the $m_F=0$ state implies that its $l_{1/2}=0$ at $t=0$. Since atoms entirely occupied $m_F = \pm1$ states, their correlation length is of the order of the system size. The situation reverses dramatically, when the~system is abruptly quenched. Therefore, the analysis of the superfluid order from the correlation length of the~$m_F=0$ spin component seems suitable for the PhOK investigation in our system. The units chosen for the characteristic length and time in the spin-1 system are~$\xi_s$~and $\tau=\hbar/(c_2 \rho)$, respectively.

\section{Scaling hypothesis and various scaling laws: numerical results}

The scaling hypothesis states that during the PhOK process a single length scale $L(t)$ is expected which increases in time as $L(t)\sim t^{1/z_d}$, where $1/z_d$ is the scaling exponent. The whole variation of the correlation function~(\ref{eq:correlation_function}) that occurs during PhOK becomes independent of time, when it is scaled by~$L(t)$. We verify the hypothesis in our system and confirm that the correlation function~$g_N^{(1)}(x, t)$ is one parameter-dependent after the proper re-scaling, i.e., $g_N^{(1)}(x, t) = f\left(x/L(t) \right)$. However, the value of the~scaling exponent $1/z_d$ strongly depends on the system parameters.

\begin{figure}[hbt!]
\centering
\includegraphics[width=0.245\textwidth]{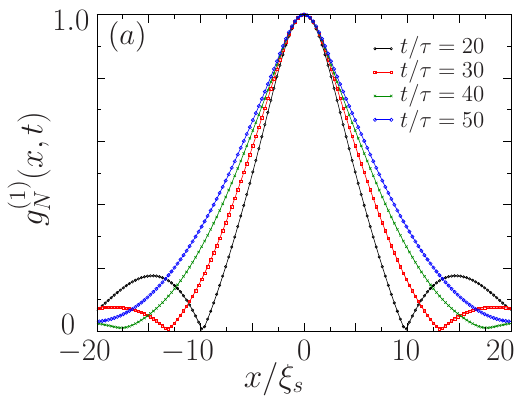}
\includegraphics[width=0.245\textwidth]{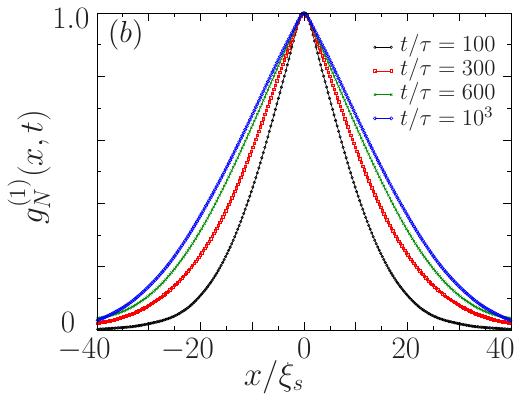}
\includegraphics[width=0.245\textwidth]{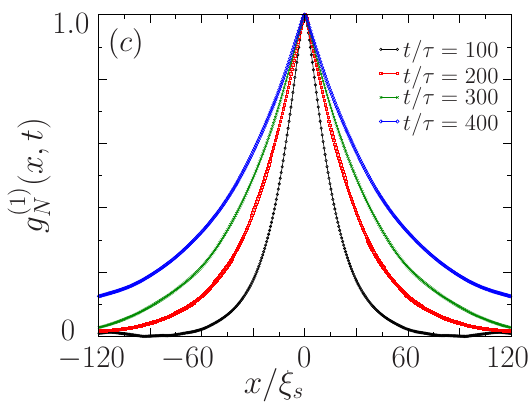}
\includegraphics[width=0.24\textwidth]{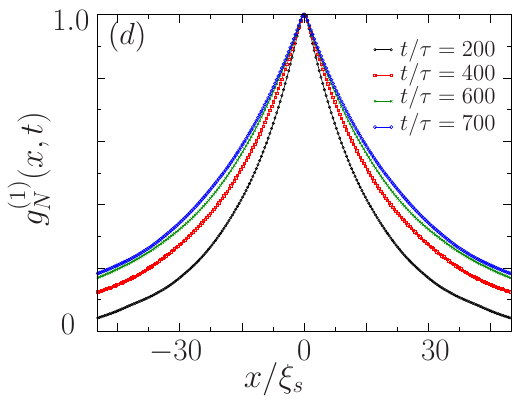}\\
\includegraphics[width=0.245\textwidth]{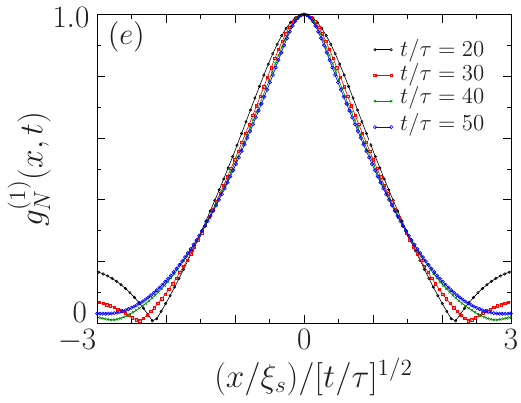}
\includegraphics[width=0.245\textwidth]{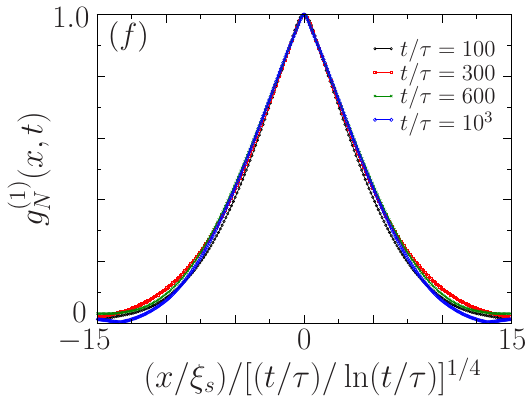}
\includegraphics[width=0.245\textwidth]{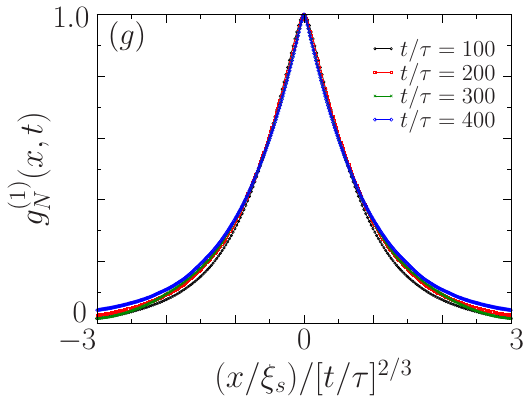}
\includegraphics[width=0.245\textwidth]{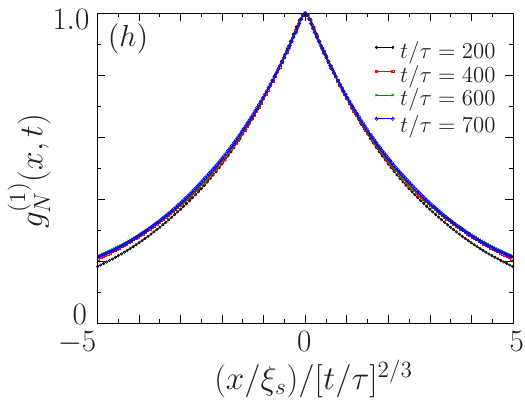}
\caption{Scaling of the correlation function $g_N^{(1)}(x, t)$ at the given moment of time as indicated in legend. Upper panels show $g_N^{(1)}(x, t)$ while bottom panels the same when lengths are re-scaled by~$L(t)$, i.e. $g_N^{(1)}(x/L(t), t)$. In (a) and (e) results for $q=0.75$\linebreak and $M/N=0.5$ are shown for short times where the scaling $L(t)\sim t^{1/2}$ is temporally observed. In (b) and (f) are the same parameters but in the long time limit when $L(t)\sim (t/{ \ln}(t))^{1/4}$ is realized. In (c) and (g) the results for $q=0.5$ and $M=0$ are shown where we recognized $L(t)\sim t^{2/3}$. Finally, in (d) and (h) data for $q=1.25$ and $M=0$ is shown where the scaling hypothesis works the best with $L(t)\sim t^{2/3}$.}
\label{fig:app1}
\end{figure}

Let us refer to the representative examples in Fig.~\ref{fig:app1}, i.e., $g_N^{(1)}(x, t)$ (upper panels) and $f\left(x/L(t) \right)$ (bottom panels) with the properly chosen the $L(t)$ scaling. Our numerical calculations show that the scaling exponent is $1/3$ and $3/2$ (or equivalently the~dynamical exponent $z_d=3$ and $z_d=3/2$) for $M=0$ and for a nonzero magnetization value when $q$ is relatively large, respectively, see an example in~Fig.~\ref{fig:app1}(c),(d) and Fig.~\ref{fig:app1}(g),(h). The two scaling exponents have already been reported for spin-1 Bose-Einstein condensates with ferromagnetic interactions characterized by negative $c_2$~\cite{PhysRevA.91.053609:,PhysRevA.95.023616}. These results can be explained on the basis of the hydrodynamic (binary mixture) model, as discussed by us in Appendix \ref{eomspin1}. In the case of macroscopic magnetization and a~relatively low value of $q$ the best matching to numerical data is observed when using scaling exponents predicted by the vector model with $L(t)\sim (t/\rm{ln}(t))^{1/4}$, see~an example in Fig.~\ref{fig:app1}(b),(f). The temporal scaling $L(t)\sim t^{1/2}$ can also be observed, see panels (a) and (e).

In the next two subsections, we discuss in details the particular scaling laws characteristic for the H and B models, and the extent of their occurrences. Our numerical findings are summarized in the last subsection in the form of the phase diagram for the~spin-1 antiferromagnetic BECs.

\subsection{In the limit of zero magnetization}

Let us start the analysis with the zero magnetization case for which the scaling laws associated with the binary mixture or H model in the Hohenberg and Halperin classification of dynamic critical phenomena can be observed.

In the model the fluid flow contributes to the transport of the~order parameter~\cite{Bray2002}, in general. This is why hydrodynamical processes like inertial or viscous growth along with diffusion mechanism are included. Each mechanism dominates at a~different stage of the domains formation, and eventually one wins at the~longest time scale. The~H model predicts the following scaling laws: $\sim t^{1/3}$ when the~diffusive transport of the~order parameter is dominant, $\sim t^{2/3}$ when the~inertia of fluids are important, and~$\sim t^{1}$ if the viscous process wins~\cite{Bray2002, PhysRevA.31.1103}. 

\begin{figure}[h!]
\centering
\includegraphics[width=0.31\linewidth]{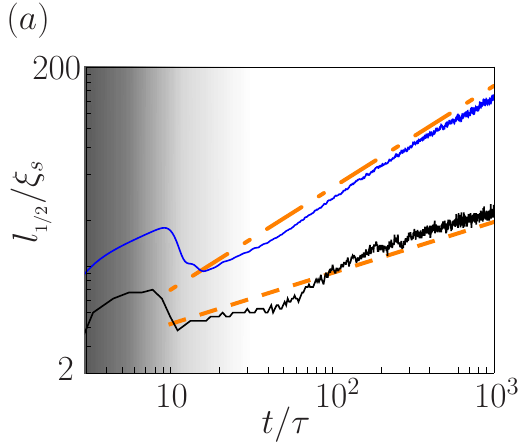}
\includegraphics[width=0.3\linewidth]{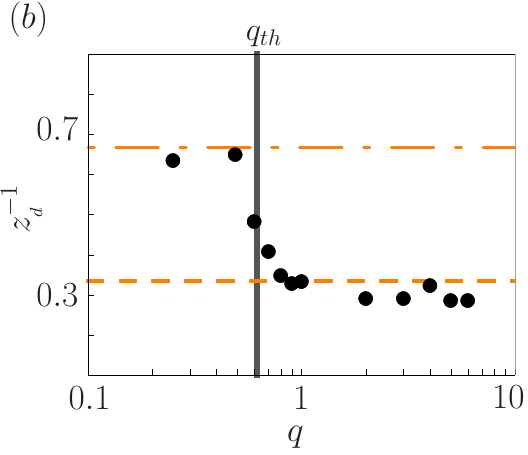}
\caption{(a) $l_{1/2}/\xi_s$ versus $t/\tau$ for $M=0$ exhibits two different scaling laws depending on the value of~$q$. Here, $q=0.5$ (blue solid line) and $q=1$ (black solid line). The scaling $\sim t^{2/3}$ is marked by the orange dot-dashed line, while $\sim t^{1/3}$ by the orange dashed line. Initial times of domains nucleation are shaded. (b) The inverse of the dynamical exponent~$z_d^{-1}$ versus $q$ is extracted at long times by fitting the function $\sim t^{1/z_d}$ to the~numerical data. The vertical thick gray line shows estimated values of the threshold point~$q_{\rm th }=0.62$.}
\label{fig:F5}
\end{figure}

Our numerical GPEs results demonstrate that the diffusive $\sim t^{1/3}$ and inertial hydrodynamic $\sim t^{2/3}$ scaling laws are revealed in the long time~limit, while the scaling law ~$\sim t^{1}$ from a pure viscous effect is absent. The exact derivation of scaling exponents from GPEs is difficult, but the estimated analysis can be performed --- a~similar one was already done on the hydrodynamic equations for the binary mixture~\cite{Bray2002}. We discuss this in Appendix \ref{eomspin1}. In general, the estimation of scaling laws for the H model requires introduction of a non-zero surface tension. In one-dimensional system described by GPEs, it equals zero. The scaling analysis we performed show that the interaction coefficient $c_0$ plays the role of surface tension and in a result, he scaling analysis of the hydrodynamic form of GPEs gives the same scaling laws as the typical H model, see Appendix \ref{eomspin1}. Consequently, the appearance of the two scaling exponents~$1/3$ and~$2/3$ in the systems described by the GPEs can be justified. 

The transition from the diffusive to the intertial hydrodynamic scaling laws is clearly visible in Fig.~\ref{fig:F5}~(b). We observe that the average domain wall width is comparable with the~width of the~phase domain itself at the transition point where the scaling exponent changes. If the size of the phase domain is larger than the size of the~average domain wall, then the diffusion transport defines the physics of the system and the scaling exponent. We~use that reasoning to estimate the transition point between $z_d^{-1}=1/3$ and $z_d^{-1}=2/3$ scaling laws. To proceed, let us assume that the fractional size of phase domain over the entire system is given by the fractional volume occupied by the~ground state phase $\rho_0$, composed of spin domains, i.e., $x_0(q)=1-\frac{\sqrt{q_1}}{\sqrt{q}}$. This~formula can be established from the~analysis of equilibrium conditions for the coexistence of 2C and $\rho_0$ phases \cite{PhysRevB.88.054508}. The~width of domain wall between these phases turns out to be set by the $q$-dependent healing length $\xi_{2C}$, estimated by expanding the Bogoliubov dispersion relation of the $m_F=0$ Zeeman component in powers of small momentum~$k$, $\epsilon_k^{(0)}=c_2 \rho \sqrt{\left(\xi_s^2 k^2 +1-q\right)^2 - \left(1-2 q\right)}\approx c_2 \rho\ q\, \left(1+ \xi_{2C}^2 k^2\right) + O(k^4)$ with $\xi_{2C}^2 = \xi_s^2 \left(1-q\right) /q^2$~\cite{PhysRevB.88.054508}. The relation  $x_0(q_{th})\approx \xi_{2C}/\xi_s$ or equivalently $1-\frac{\sqrt{q_1}}{\sqrt{q_{th}}}\approx \sqrt{1-q_{th}} /q_{th}$, gives the desired condition for the transition point $q_{th}$ between the~two different scaling laws. Whenever $q>q_{th}$, the diffusive transport governs domains coarsening. The resulting estimate for~$q_{th}$ is shown in Fig.~\ref{fig:F5}(b) by the~vertical solid line, and in Fig.~\ref{fig:F6} by the dashed line.

At this juncture, it is useful to connect the appearance of domains of phase (PhOK) and also domains in the real space, to the occupations of atoms in particular Zeeman states. The system we consider is three component in general. When $M$ decreases at relatively low~$q$ the $m_F=0$ component becomes macroscopically occupied at~longer time scales, while the remaining two components $m_F=\pm 1$ are occupied marginally, both to almost the same extent. For the purposes of the~dynamics one can expect that the~properties of atoms remaining in the $m_F=\pm 1$ components become identical. The~system starts then to behave like a binary mixture composed of two species of atoms: these in the~$m_F=0$ state and those in $m_F=\pm 1$ treated together as the~second species. Similarly, in the large $q$ limit occupation of the $m_F=-1$ component becomes marginal while the remaining two $m_F=0,1$ are of the only importance. In both regimes of parameters, the binary mixture description of PhOK could become suitable for our system, and therefore, the resulting scaling exponents could appear. 

\subsection{Macroscopic magnetization}

\begin{figure}[h!]
\centering
\includegraphics[width=0.245\textwidth]{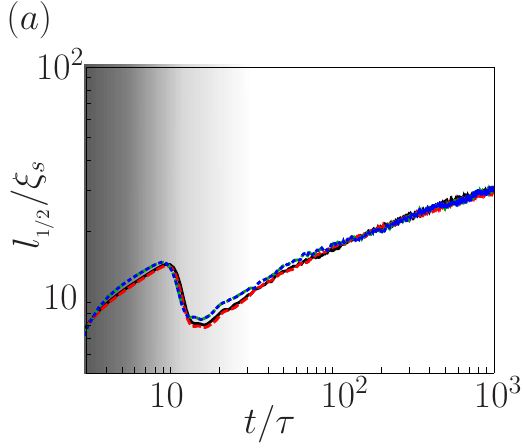}
\includegraphics[width=0.245\textwidth]{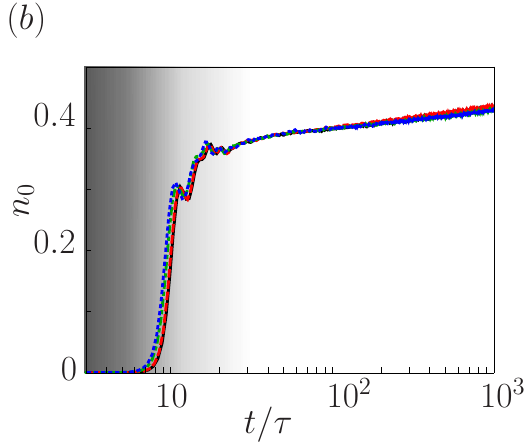}
\includegraphics[width=0.245\textwidth]{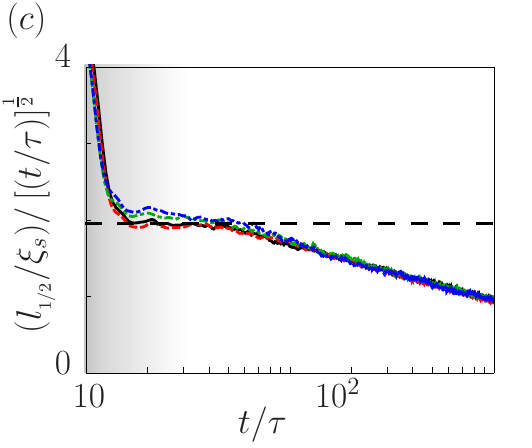}
\includegraphics[width=0.245\textwidth]{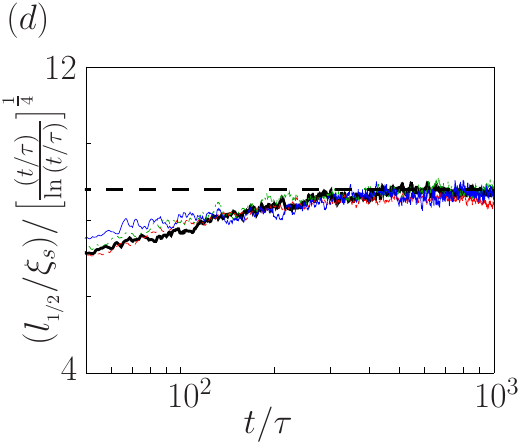}
\caption{(a) Scaled correlation length $l_{1/2}/\xi_s$ versus scaled time $t/\tau$ given for the linear densities $\rho=14.3\mu{\rm m}^{-1}$ (black solid line), $\rho=10\mu{\rm m}^{-1}$ (red dashed line), $\rho=2\mu{\rm m}^{-1}$ (green dot-dashed line), $\rho=1\mu{\rm m}^{-1}$ (blue dot-dot-dashed line) in the effective vector field regime for $N=10^4$, $M=N/2$ and $q=0.75$. (b) Variation of $n_0$ in time. Regime of scaling laws referencing model A with a temporarily non conserved order parameter (c), and model B with a conserved order parameter at long times (d). The~horizontal dashed line is added to guide the eye. Initial times of domains nucleation are shaded. Note, the~linear scale on the vertical axis in (c) and (d).}
\label{fig:F3}
\end{figure}

In the case of macroscopic magnetization~$M$, we can talk about the regime of parameters for which occupations of all three Zeeman components are significant. Also then the variation of the scaled correlation length $l_{1/2}/\xi_s$ versus scaled time $t/\tau$ is expected, as shown in~Fig.~\ref{fig:F3} for various system densities~$\rho$. This time, an~increase of the correlation length does not exhibit the scaling attached to the binary mixture models. Instead, we observe that $l_{1/2}$ reveals growth typical for vector field models~\cite{Bray2002}. Those models predict $L(t)\sim t^{1/2}$ (model~A) and $L(t)\sim (t/{\rm ln} t)^{1/4}$ (model~B) for non conserved and conserved order parameters, respectively.

Better understanding of the evolution of the order parameter in the case of considered system can be provided with definition of $n_0=N_0/N$ --- the fractional number of atoms in the $m_F=0$ Zeeman component, where $N_0=\int {\rm d}x\, |\psi_0(x)|^2$.\linebreak It~is shown in~Fig.~3(b). Notice that for early times $n_0$ is susceptible to large fluctuations. In~the~range $\sim 10 \tau \dots 40\tau$, the~order parameter is found to be evidently non conserved. The correlation length scaling exponent is $1/2$ there, see~Fig.~3(c). However, for~the~longest time scale which we are interested in, the change of $n_0$ is less significant although its variance is far from being equal to zero. Despite that fact, the~value of the scaling exponent changes to $1/4$ with logarithmic correction. The two scaling laws can also be observed by properly re-scaling the~correlation function $g_N^{(1)}(x, t)$ as demonstrated in Fig.~\ref{fig:app1}. Note, however, that the universal character can have only the scaling revealed in the long time limit, that is $z_d^{-1}=1/4$. The temporal $1/2$ scaling is mentioned by us only to demonstrate the multiscaling behaviour of PhOK characteristic for our system in early times. The similar observation and conclusion was also made for ferromagnetic spin-1 Bose-Einstein condensates~\cite{PhysRevLett.119.255301}. 

The~character of the system smoothly transforms to the binary mixture when the~value of $q$ increases. The occupation of the~$m_F=-1$ Zeeman component diminishes to zero and the remaining two, $m_F=0,1$ are of the only importance. The change of the scaling exponent to $1/3$ typical for the H model is observed by us independently of $M$. 

\subsection{Classification of scaling laws}

\begin{figure}[h!]
\centering
\includegraphics[width=0.245\textwidth]{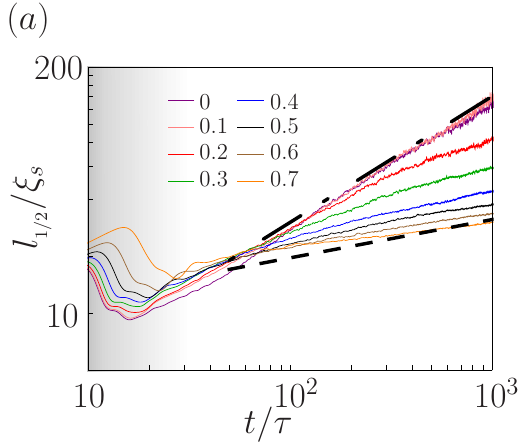}
\includegraphics[width=0.245\textwidth]{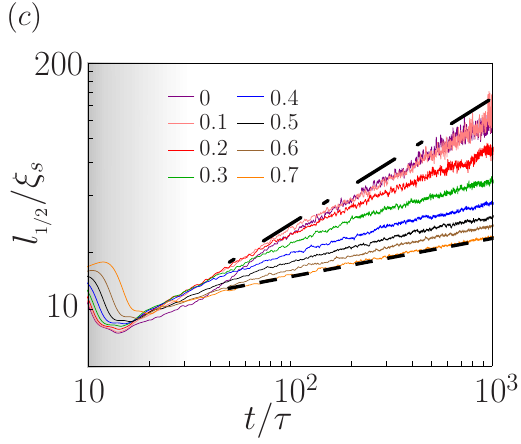}
\includegraphics[width=0.245\textwidth]{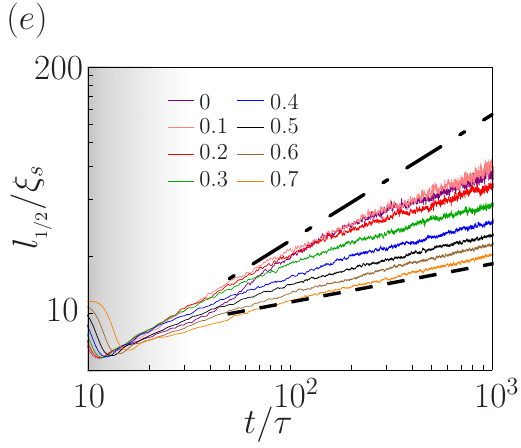}
\includegraphics[width=0.245\textwidth]{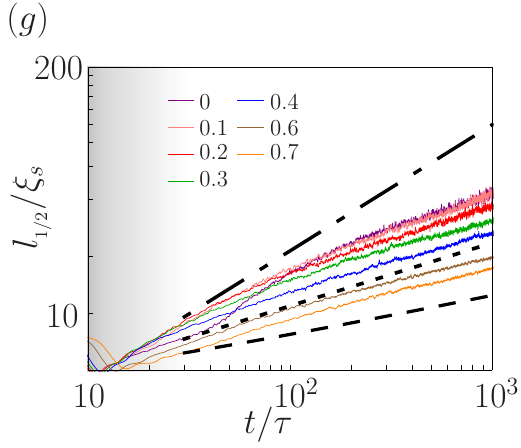} \\
\includegraphics[width=0.245\textwidth]{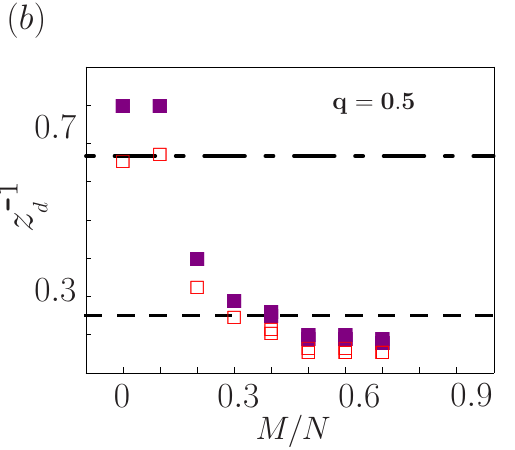} 
\includegraphics[width=0.245\textwidth]{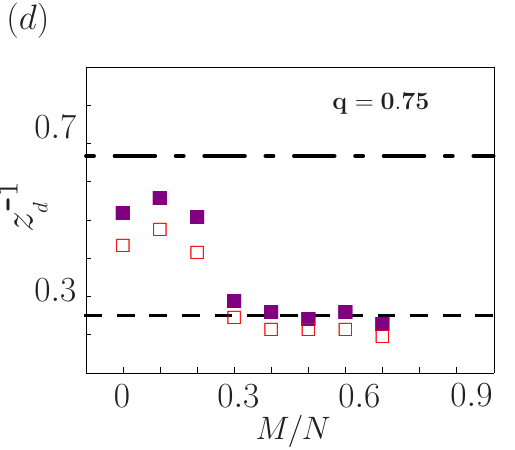}
\includegraphics[width=0.245\textwidth]{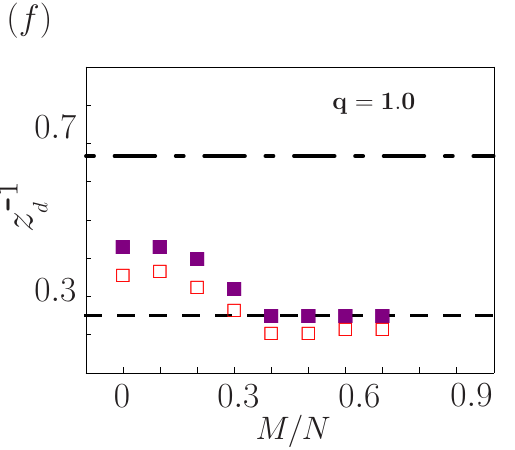}
\includegraphics[width=0.245\textwidth]{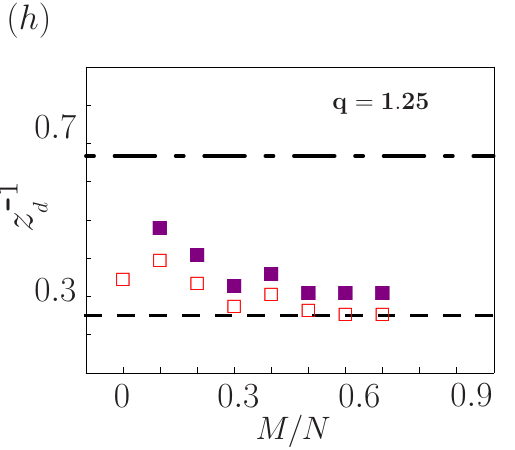}
\caption{Upper panels: $l_{1/2}/\xi_s$ versus $t/\tau$ for (a) $q=0.50$, (c) $q=0.75$, (e) $q=1.00$, (g)~$q=1.25$, and the fractional magnetization $M/N$ values given in the legend. The dashed line indicates the scaling $\sim (t/\ln{t})^{1/4}$ resulting from the vector model B, while the respective dot-dashed and dotted line indicate the scaling $\sim t^{2/3}$ and $\sim t^{1/3}$ typical for the hydrodynamic model~H. Bottom panels: The~inverse of the dynamical exponent~$z_d^{-1}$ extracted at long times from fitting the data shown in the corresponding upper panel. Open points demonstrate results by fitting the function $\sim t^{1/z_d}$ to the~numerical data, while the closed one by fitting the function $\sim (t/\ln(t))^{1/z_d}$. }
\label{fig:F4}
\end{figure}

To summarize the results at this stage: we show the~variation of scaling laws in time for $q=0.5$, $0.75$, $1.0$, $ 1.25$ in~Fig.~\ref{fig:F4} depending on the magnetization value. Our observations confirm the presence of mechanisms which are competitive with each other and they lead finally to different behaviour in the long times limit. Therefore, several values of the~growth exponents can emerge during the~later evolution, namely $z_d\to 3/2$, $z_d\to 3$ or even to $z_d \to 4$ with logarithmic correction. The change of scaling exponent while varying magnetization is typical for our system and appears in a wide range of the value of $q$.

Careful analysis of the numerical results indicates an important role of occupations of particular Zeeman states. We~conclude that occupations of atoms in particular Zeeman components give an intuitive picture of which of the scaling law will be revealed in the long dynamics of our system. The~vector (B) model fits the best when occupations of all the three Zeeman components are significant. It takes the place for macroscopic magnetization for values of $q$ a bit larger than the critical one,~$q_c$. In~turn, a~high concentration of atoms can stimulate stronger interactions, the liquid character and thus the hydrodynamic description would be more accurate. It is when magnetization is small and in the large $q$ limit, $q\gg q_c$. In this regime, the hydrodynamic (H) model seems to match the~best. 

\begin{figure}[hbt!]
\centering
\begin{picture}(0,100)
\put(-215, -4){\includegraphics[height=3.4cm]{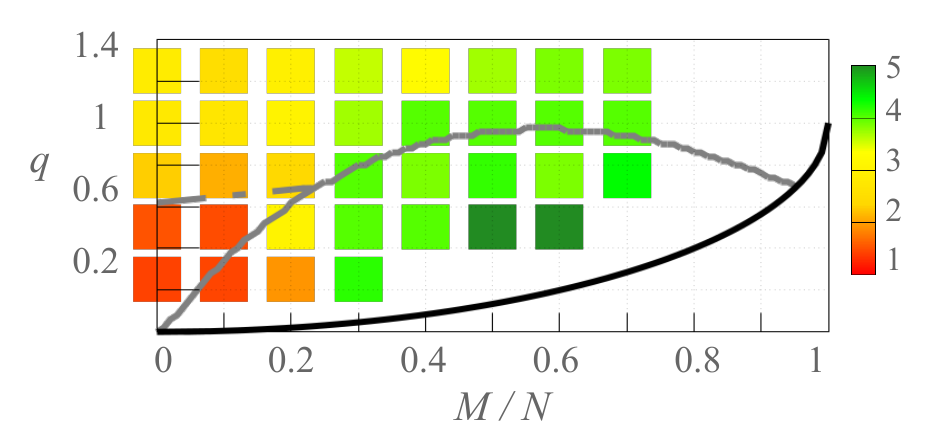}} 
\put(-230, 90){\small $(a)$} 
\put(5, -1){\includegraphics[height=3.2cm]{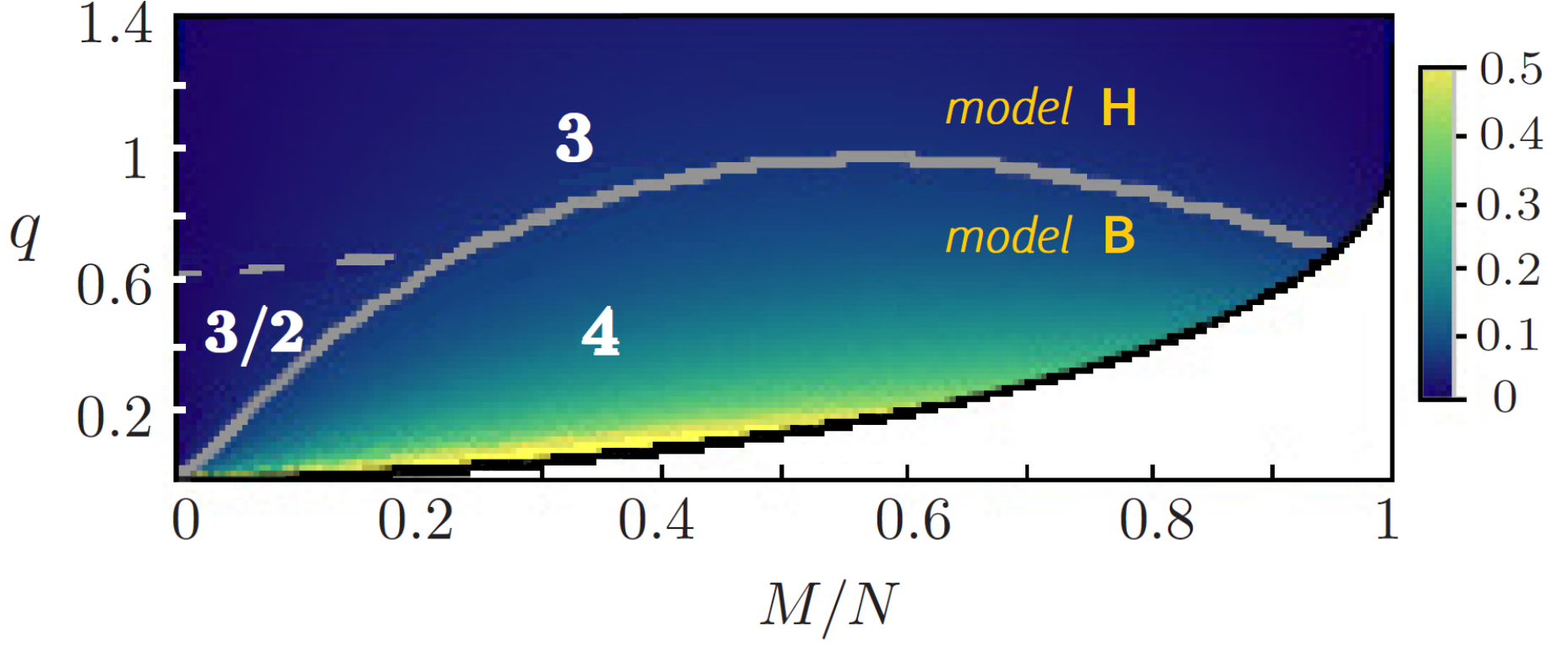}}
\put(-10, 90){\small $(b)$} 
\end{picture}
\caption{ The values of dynamical exponent $z_d$ versus $M/N$ and $q$ revealed in the antiferromagnetic condensate, marked by color in (a) while in (b) by white numbers. (a) The values of colored boxes are numerically fitted results from Fig.~\ref{fig:F5} and Fig.~\ref{fig:F4}, provided that below the solid gray line logarithmic correction to the~scaling law is taken into account. (b) When the occupation $N_{-1}/N$ (shown by color) is substantial or marginal, the B or H model's scaling is realized, respectively. The~solid gray line marks the border between the H and B models estimated by the~condition $N_{-1}/N =0.025$. The~gray dashed line is the transition point $q_{th}$ where the two different scaling exponents apply for the model~H. The~top edge of the white area, which is restricted by relation $q_c=1-\sqrt{1-(M/N)^2}$, determines where the~critical transition between $2C$ and $2C+\rho_0$ phases takes place. }
\label{fig:F6}
\end{figure}

The~border between both models can be deduced by matching when the~number of atoms in the~$m_F=-1$ component vanishes, i.e., $N_{-1}(M, q)/N \to 0$. This is illustrated in~Fig.~\ref{fig:F6}. The~vector field model is realized below the~gray solid border line, while the~binary mixture model applies above it. The~change between the two regions is smooth, and so $z_d$ does not perfectly reflect the particular model all around the border line.

\section{Discussion and conclusion}

In this paper, we have explored the superfluid phase-ordering dynamics of an antiferromagnetic spin-1 condensate quenched from the antiferromagnetic state to a state where domains of atoms with different spin projections separate. We have found that the growth of domains is scale-invariant with various dynamic critical exponent $z_d$ that are typical for the B and H models. We~classified the various scaling exponents due to the system parameters. 

When the occupation of the $m_F=-1$ Zeeman state is significant, We conclude that the scaling exponents typical for vectors fields model appear ($z_d =4$ with logarithmic correction). At short times we observe also that temporally the correlation length $l_{1/2}$ exhibits the scaling $L(t)\sim t^{1/2}$. These all properties are characteristic for multiscaling of the PhOK.

On the other hand, when the occupation of the $m_F=-1$ component is marginal the scaling laws typical for the H model are realized, namely $z_d=3/2$ when $q$ and magnetization are small, and $z_d=3$ in the large $q$ limit for any $M$. While the appearance of scaling laws typical for the H model can be justified from the hydrodynamic form of GPEs, the occurrence of scaling typical for B model is not evident. The latter provides an interesting direction for future work.

To conclude: the antiferromagnetic spin-1 condensate captures the~universal two-model feature of the PhOK in the long time limit. The system parameters \mbox{($q$, $M/N$)} set the physics, and determine the dynamical scaling exponent corresponding to the model H or B. Switching between the models by changing the initial state or the $q$ parameter is allowed. 

\section*{Acknowledgements}
We wish to thank Piotr Deuar and Micha\l{} Matuszewski for useful discussions, and Oleh Hul for careful reading of manuscript.
EW and JP acknowledge support from the National Science Centre (Poland) grants Nos. 2015/18/E/ST2/00760 and 2012/07/E/ST2/01389, respectively.

\section*{Author contributions statement}
J.P. and E.W. equally contributed to the numerical calculations, analysis of the results and writing the manuscript. A.S. contributed to the numerical simulations of the $M/N=0.5$ case.


\begin{appendix}

\section{\textcolor{col}{Scaling analysis from spin-1 hydrodynamic equations}}
\label{eomspin1}

\textcolor{col}{
In the following, we present the hydrodynamic forms of the GPEs. The wave functions in Eq.~(\ref{eq:GPEs}) describing atoms in particular components are replaced by $\psi_{m_{F}}=\sqrt{n_{m_F}}\, e^{i\, \theta_{m_F}}$ -- according to Madelung transformation. After some algebra, we obtain corresponding equations for the densities $n_{m_F}$, namely}
\textcolor{col}{\begin{equation}
    \dot{n}_{m_F}+\frac{\hbar}{m}v_{m_F}\nabla n_{m_F} 
    = \frac{\hbar}{m} n_{m_F} \nabla v_{m_F} + 2 \frac{c_2}{\hbar} (|m_F| - 2)\, n_0\, \sqrt{n_1 n_{-1}} \sin(\theta)
\end{equation}
and the corresponding Navier-Stokes equations for velocities
$v_{m_F}=\frac{\hbar}{m}\nabla \theta_{m_F}$
\begin{equation}\label{eq:NS-spin-1}
    \dot{v}_{m_F} + 4 \frac{m}{\hbar} {v}_{m_F} \nabla {v}_{m_F} 
    = 2 \frac{m}{\hbar} \nabla \left( \frac{\nabla^2 \sqrt{n_{m_F}}}{\sqrt{n_{m_F}}} \right)  + \nabla U_{m_F}  
    - c_2 (2 - |m_F|) \nabla \left( n_0 \sqrt{\frac{n_1 n_{-1} }{n_{m_F}}} \cos(\theta)\right),
\end{equation}
where $\theta = \theta_1 + \theta_{-1} - 2 \theta_0$, and $U_1 =  c_0 n + q c_2 \rho + c_2 (n_1 - n_{-1} + n_0)$, $U_0 = c_0 n + c_2 (n_1 + n_{-1})$, $U_{-1} = c_0 n + q\, c_2 \rho + c_2 (n_{-1} - n_{1} + n_0)$.}

We will perform here the scaling analysis corresponding to the one presented in~\cite{Bray2002} for the hydrodynamic (B) model. The chemical potential can be identified as $\tilde{\mu}=U_0$, see details in ~\cite{PhysRevB.88.054508}, and therefore the derivative of chemical potential as $\phi \nabla \tilde{\mu} \to \nabla U_{0}$. 
Taking this as an assumption one can perform the~scaling analysis for $m_F=0$ as in~below.\\
{\large \textcircled{\small 1} } 
Scaling by diffusion mechanism:\\
In the diffusive regime it is assumed that the rate of change of the domain size $\frac{dL}{dt}$ is associated with the~chemical potential gradient $\nabla U_{0}$. The scaling of $U_{0}$ is linked to the~scaling of the local density $U_0 \sim c_0 n \sim c_0N/L$, therefore one has:
\begin{equation}
 \frac{dL}{dt} \sim \frac{c_0}{ L^2} \qquad \Longrightarrow \quad  L(t) \sim t^{1/3}.
 \label{gameq}
\end{equation}
Note, the role of the surface tension $\sigma$ typically introduced in analysis of binary mixture~\cite{Bray2002} is played here by the interaction coefficient $c_0$.\\
{\large \textcircled{\small 2} } 
Scaling by inertial growth: \\
If the inertial terms become important, the scaling law is given by the relation \mbox{$(v_{0} \nabla) v_{0} \sim \nabla U_{0}$.} Then, one recovers the same scaling as for the H model, namely
\begin{equation}
\left(\frac{dL}{dt}\right)^2 \sim \frac{c_0}{L}  \quad \Longrightarrow   \quad
   L(t)\sim  t^{2/3}.
\label{inereq}
\end{equation}

\end{appendix}

\bibliography{biblio.bib}
\end{document}